\documentclass[referee]{raa}            

\usepackage{graphicx,times}             
\usepackage{natbib}
\usepackage{amssymb,amsmath}
\bibpunct{(}{)}{;}{a}{}{,}

\usepackage[a4paper=true,dvipdfm=true,pagebackref=true]{hyperref}
\hypersetup{colorlinks = true, linkcolor = green, anchorcolor = red, citecolor = blue, filecolor = red, pagecolor = red, urlcolor = red}

\newcommand{\sgra}{Sgr~A$^*$}		
\newcommand{\kms}{\,km\,s$^{-1}$}	
\newcommand{\kmskpc}{\,km\,s$^{-1}$\,kpc$^{-1}$}
\newcommand{\masyr}{\,mas\,yr$^{-1}$}	

\begin{document}

   \title{On the possible orbital motion of Sgr A* in the smooth potential of~the Milky Way
}

   \volnopage{Vol.0 (20xx) No.0, 000--000}      
   \setcounter{page}{1}          

   \author{Igor' I. Nikiforov
   \and Angelina V. Veselova
   }

   \institute{St. Petersburg State University, Universitetskij Prospekt 28, Staryj Peterhof,
St. Petersburg 198504, Russia; {\it i.nikiforov@spbu.ru (IIN); linav93@yandex.ru (AVV)}\\
\vs\no
   {\small Received~~20xx month day; accepted~~20xx~~month day}}

\abstract{The modern accuracy of measurements allows the residual/peculiar
(Galactocentric) velocity of the supermassive black hole (SMBH) in our Galaxy, \sgra, on
the order of several kilometers per second. We integrate possible orbits of the SMBH
along with the surrounding nuclear star cluster (NSC) for a barred model of the Galaxy
using modern constraints on the components of the SMBH Galactocentric velocity. Is is
shown that  the
range of oscillations of the SMBH\,+\,NSC in a regular Galactic field in the plane of the
Galaxy allowed by these constraints strongly depends on the set of central components of
the Galactic potential.  If the central components are represented only by a bulge/bar, for a point
estimate of the SMBH Galactocentric velocity, the oscillation amplitude does not exceed
7\,pc in the case that
a classical bulge is present  and reaches 25\,pc if there is no
bulge; with SMBH velocity components within the $2\sigma$ significance level, the
amplitude can reach 15 and 50\,pc, respectively. However, when taking into account
the nuclear stellar disk (NSD), even in the absence of a bulge, the oscillation amplitude
is only 5\,pc for the point estimate of the SMBH velocity, and 10\,pc for the $2\sigma$
significance level. Thus, the possible oscillations of the SMBH\,+\,NSC complex from the
confirmed components of the Galaxy's potential are mostly limited by the NSD, and even
taking into account the uncertainty of the mass of the latter, the oscillation amplitude
can hardly exceed $13\,\text{pc}=6'$. 
\keywords{Galaxy: centre --- Galaxy: fundamental parameters --- Galaxy: kinematics and dynamics}
}

   \authorrunning{I.I. Nikiforov \& A.V. Veselova}            
   \titlerunning{On the possible orbital motion of Sgr A* in the smooth potential of~the Milky Way}  

   \maketitle

%
%
\section{Introduction}           
\label{sect:intro}

Modeling the orbital motions of S stars in the vicinity of the supermassive black hole
(SMBH) \sgra\  in the central region of the Milky Way allows accurate estimates of the
distance $R(\text{BH})$ from the Sun to this object (see, e.g., \citealt{B-HG16,dGB16}).
The most reliable are the results of recent analyses of data on the star S2 (S0-2): 
\cite{GC19} reported $R(\text{BH})=8178\pm 13_\text{\,stat.}\pm
22_\text{\,sys.}$\,pc; \cite{Dea19} measured $R(\text{BH})=7946 \pm 50\,(\text{stat.}) \pm
32\,(\text{sys.})$\,pc (both estimates assume General Relativity is true).  It is commonly
believed that the SMBH is exactly at the (bary)center of the Galaxy (e.g.,
\citealt{dGB16}). Then $R(\text{BH})$ measurements can be considered as estimates of the
distance $R_0$ to the center of the Galaxy: $R_0=R(\text{BH})$. Moreover, such estimates
are absolute (i.e., not using luminosity calibrations; see the classification in
\citealt{N04}) and have high (at least formal) precision and accuracy.

At the moment, the $R_0$ estimates obtained by the orbit method by the two research groups
differ significantly (at the $3.6\sigma$ level for the above estimates, taking into
account the systematic uncertainties specified by the authors). Yet $R(\text{BH})$
measurements already have high precision, which will grow in the future as data
accumulate. It makes sense to consider the scale of the possible deviation of the SMBH
from the barycenter of the Galaxy, which may not be negligible compared to the current and
future accuracy of  $R(\text{BH})$ estimates. \cite{B94} pointed to the possibility of
oscillations of the \sgra\ system and other central Galactic mass concentrations in a
``fairly shallow'' bar potential. Some explorations of the dynamics of SMBHs in galactic
cores show that the displacement of the SMBH from the geometric center of the galaxy due
to interaction with globular clusters and stars can reach several parsecs
(\citealt{KO08,DiCea19}). In the case of M87, the SMBH appears to be off-centered in the
parent galaxy by $6.8\pm 0.8$\,pc (\citealt{Bea10}).

The assumption that the SMBH is essentially at rest at the Galactic center is based
largely on the marginally nonzero peculiar proper motion of \sgra\ (e.g.,
\citealt{RB04,B-HG16}). However the accuracy of modern measurements does not exclude the
residual velocity of \sgra\ in Galactic longitude and the peculiar radial velocity of the
SMBH on the order of several kilometers per second (see Sect.~\ref{sect:velocity}).
Further in this paper, for a barred model of the Galaxy (Sect.~\ref{sect:potential}), we
determine possible orbits of the SMBH, along with its host nuclear star cluster
(NSC), using modern constraints on the components of the SMBH residual/peculiar velocity
to estimate the scale of possible SMBH\,+\,NSC oscillations relative to the barycenter
(Sect.~\ref{sect:orbits}) and then we discuss the results
(Sect.~\ref{sect:discussion}).

\section{Measurements of the peculiar/residual velocity of the SMBH (Sgr~A*)}
\label{sect:velocity}


In this work, we consider the SMBH
and the NSC as a single complex with a common orbit in a regular Galactic field.
Indeed, the significant mass ${\sim}2\times10^7M_{\odot}$ 
and compactness (the effective radius is ${\sim}5$\,pc) of the NSC (e.g.,
\citealt{B-HG16,G-Cea20}) should greatly
limit the possible deviations of the black hole from its center (we test this in
Sect.~\ref{sect:discussion}). The detection of a stellar cusp around the SMBH (e.g.,
\citealt{G-Cea20}) indicates a partial relaxation of the
NSC (e.g., \citealt{Baumgardt+18}). 
In the projection, the SMBH is observed near the center of the NSC (e.g.,
\citealt{Feldmeier,G-Cea20}). 
Therefore, considering the SMBH and NSC as a single complex in which the SMBH is located
near its barycenter seems acceptable, at least in the first approximation. The SMBH\,+\,NSC
complex itself can undergo oscillations in the potential of more extended Galactic
components, and we aim to find out the range of these oscillations depending on the
composition of the potential model.

A number of research results suggest that the NSC is not fully relaxed
and its populations of different ages are not fully mixed (e.g.,
\citealt{B-HG16,G-Cea20}). A manifestation of this may be the asymmetry
of the NSC's rotation curve in the sense that the absolute velocities are higher on the
eastern side than on the western side ( \citealt{Feldmeier}). Restoring the symmetry
by assuming a non-zero net radial velocity of the NSC, we obtain a formal estimate of it,
$V_r^\text{LSR}(\text{NSC})=+10.6 \pm 1.9$\,km\,s$^{-1}$,  which strongly disagrees with
recent accurate measurements of the SMBH's radial velocity (see Sect.~\ref{sect:radial
velocity}) and can hardly be attributed to the barycenter of the SMBH\,+\,NSC. Therefore, as
estimates of the velocity components of the SMBH\,+\,NSC complex, we use the estimates
obtained for the SMBH, which are discussed below.

Further, by default, under the orbits of the SMBH and other terms describing its
possible motion, we will understand the corresponding terms related to the SMBH\,+\,NSC
complex.

\subsection{Radial velocity of the SMBH relative to the LSR}
\label{sect:radial velocity}

The peculiar radial velocity of the SMBH, $V_r^\text{LSR}(\text{BH})$, is determined by
the orbit method (\citealt{GC19,Dea19}). The heliocentric radial velocity of the SMBH is
\begin{equation}
 V_r(\text{BH})=V_r^\text{LSR}(\text{BH})-u_\odot^\text{LSR},
\end{equation}
where $u_\odot^\text{LSR}$ is the solar (peculiar) velocity with respect to the Local
Standard of Rest (LSR) towards the Galactic center. Then
\begin{equation}
 V_r^\text{LSR}(\text{BH})=V_r(\text{BH})+u_\odot^\text{LSR}.
\label{eq:VrLSR}
\end{equation}
\cite{GC19} give an estimate of $V_r^\text{LSR}(\text{BH})=-3.0\pm 1.5$\,km\,s$^{-1}$
with $u_\odot=11.10^{+0.69}_{-0.75}$\,km\,s$^{-1}$ from \cite{Schea10}. If instead of this
separate $u_\odot$ estimate we use the summary value of $u_\odot=10.0\pm 1$\,km\,s$^{-1}$
from the review by \cite{B-HG16}, then according to Equation~\eqref{eq:VrLSR} we get
$V_r^\text{LSR}(\text{BH})=-4.1\pm 1.5$\,km\,s$^{-1}$. The last estimate shows that we
cannot exclude values of $-1.1 \le V_r^\text{LSR}(\text{BH})\le -7.1$\,km\,s$^{-1}$ at the 2$\sigma$
level. Therefore, as the initial radial velocity of the SMBH for the integration of its orbit,
the values of $V_r^\text{LSR}(\text{BH})=-4.1$\,km\,s$^{-1}$ (hereinafter the nominal
value) and $V_r^\text{LSR}(\text{BH})=-7.1$\,km\,s$^{-1}$  (hereinafter the 2$\sigma$-value) were
taken. Less accurate values of $V_r^\text{LSR}(\text{BH})=(-3.6, -6.2)\pm 3.7_\text{\,stat.}\pm
0.79_\text{\,sys.}$\,km\,s$^{-1}$ reported by \cite{Dea19} are within these limits.

Note that the  velocity $V_r^\text{LSR}(\text{BH})$ from \cite{GC19}, although small, is
marginally significantly (${\ge}2\sigma$) different from zero for both $u_\odot$ values.

\subsection{Longitude residual velocity of the SMBH}
\label{sect:longitude velocity}

The apparent motion of \sgra\ in Galactic longitude is $\mu_l(\text{BH})=-6.379 \pm
0.026\text{\masyr}$ \citep{RB04}, which translates to $\mu_l(\text{BH})=-30.24 \pm
0.12$\kmskpc. The value of $\mu_l(\text{BH})$ can be written as
\begin{equation}
 \mu_l(\text{BH})=\mu_l^0(\text{BH})-\omega_{\sun},\qquad
 \omega_{\sun}=\omega_0+v_{\sun}/R_0,
\end{equation}
where $\omega_{\sun}$ is the solar angular rotation rate; $\omega_0$ is the Galactic
angular rotation rate of the considered (on non-local scales) Galactic subsystem at the
Sun, i.e., the rate of the reference system, which can be called the {\em non-local
standard of rest of objects\/}; $v_{\sun}$ is the solar (residual) motion in the direction
of Galactic rotation relative to this non-local standard of rest; $\mu_l^0(\text{BH})$ is
the residual motion of \sgra\ in $l$. A value of $\mu_l(\text{BH})$ makes it possible to
estimate the linear residual motion of \sgra\ in $l$, using relation
\begin{equation}
 V_l^0(\text{BH})\equiv\mu_l^0(\text{BH})R_0=[\mu_l(\text{BH})+\omega_{\sun}]R_0,
\end{equation}
i.e., without assumptions about the solar peculiar velocity, since values of
$\omega_{\sun}$, $\omega_0$, $v_{\sun}$, and $R_0$ can be directly determined from the
analysis of kinematics of a Galactic subsystem. Values of $\omega_{\sun}$ calculated by us
from kinematic parameters found by \cite{Ret17} from data on Galactic masers are in the
range from  $30.72\pm {\le}0.47$\kmskpc\ to $31.16\pm {\le}0.54$\kmskpc\ for different
kinematic models, which corresponds to $V_l^0(\text{BH})=
+3.8\pm {\le}3.9$ and $+7.4\pm {\le}4.4$\kms, respectively. 
To estimate the largest range of SMBH oscillations, for the orbit integration we took the
second of these values as the nominal initial SMBH velocity in longitude
$V_l^0(\text{BH})=+7$\kms; correspondingly, the 2$\sigma$-value is
$V_l^0(\text{BH})=+16$\kms.

\section{Model potential of the Galaxy}
\label{sect:potential}

Orbits were integrated using a gravitational potential, which consists in general of five
components: the Galactic disk and the nuclear stellar disk~(NSD) are modeled by the
Miyamoto~\& Nagai potentials, the halo by a logarithmic potential, the bar by a Ferrers
potential of an inhomogeneous triaxial ellipsoid (with parameter $n = 2$), and the bulge by
three different potentials. For all components except the bulge, we used the same
expressions and parameter values (in particular, $R_0=8$\,kpc) as in
\cite{C-Dea13}. These authors presented the following models of potentials.

The Galactic disk potential was set by the Miyamoto~\& Nagai model
 \begin{equation}
 \label{MNmodel}
\Phi(R,z) = - \frac{GM_{\text{comp}}}{\sqrt{R^2 + \left(a  + \sqrt{z^2+b^2} \right)^2}}\,,
\end{equation}
where $M_{\text{comp}}$ is the component's mass, with parameter values of $a =
6.5$\,kpc, $b = 0.26$\,kpc, and $M_{\text{comp}}=M_{\text{disk}} = 1.1\times10^{11} M_{\odot}$. 

The logarithmic potential of the halo had the form 
\begin{equation}
\Phi(r) = v_{\text{halo}}^2\ln(r^2+d^2),
\end{equation}
where $v_{\text{halo}} = 121.9\,\text{km\,s}^{-1}$ and $d = 12$~kpc.

To model the bar's potential the Ferrers potential was used (see Binney \& Tremaine
2008). The density distribution in the bar was as follows
\begin{equation}
\rho(m^2) = \begin{cases} \rho_0\left(1 - \frac{m^2}{a_1^2}\right)^2 &\text{for}~m \le a_1\,, \\
0 & \text{for}~m > a_1\,,
\end{cases} \quad \text{where} \quad m \equiv a_1^2\sum\limits_{i=1}^3 \frac{x_i^2}{a_i^2}\,.
\end{equation}
Here $a_1,a_2,a_3$ are the semi-axes of the triaxial ellipsoid and $x_1,x_2,x_3$ are
the Cartesian coordinates in the system of the rotating bar. The potential of the bar is
\begin{equation}
\Phi(x_1,x_2,x_3) = -\frac{\pi G\rho_0a_1a_2a_3}{3} \int\limits_0^{\infty}\frac{d\tau}{\sqrt{(\tau+a_1^2)(\tau+a_2^2)(\tau+a_3^2)}}\times\left( 1- \sum\limits_{i=1}^3\frac{x_i^2}{\tau+a_i^2}\right)^3.
\end{equation}
\cite{C-Dea13} gave the following values of the semi-axes of the bar: $a_1 = 3.14$\,kpc,
$a_2 = 1.178$\,kpc and $a_3 = 0.81$\,kpc. The central density $\rho_0$ is obtained using the
total mass of the bar and its volume.

The angular velocity of rotation of the bar is assumed to be 
$\omega_\text{bar}=40$\kmskpc, and the angle of its inclination (the Galactocentric
longitude of the bar's edge nearest to the Sun, measured from the direction of the Sun
clockwise) is $\varphi_0 = 25\dg$ (\citealt{B-HG16}).

A Hernquist potential, applied by \cite{C-Dea13} to represent the bulge, is unsuitable for
the purposes of this study, since for this model the force as a function of coordinates
has a singularity at the center (a nonzero value). Therefore, we have considered three
other options for modeling the potential of the bulge.

(i) The Miyamoto~\& Nagai potential~\eqref{MNmodel}
with parameter values of $a = 0.04$\,kpc and $b = 0.2$\,kpc for $R_0 = 8.5$\,kpc
(\citealt{Nk92}) multiplied by the correction coefficient $8/8.5$.

(ii) The isochrone potential (\citealt{BT08})
\begin{equation}
\Phi(r) = - \frac{GM_{\text{bulge}}}{b_1+\sqrt{b_1^2 + r^2}}\,,
\end{equation}
with $b_1=0.15$\,kpc to get in some sense an intermediate variant between models (i) and (iii).

(iii) The Plummer model 
\begin{equation}
\Phi(r) = -\frac{GM_{\text{b}}}{\sqrt{r^2+c^2}}\,,
\end{equation}
where $c = 0.3$\,kpc (\citealt{KO08}).

With accepted parameters, the Miyamoto~\& Nagai potential is deepest at the center, and
the Plummer model has the lowest peak radial force.

At the moment it is not known for certain whether a classic bulge exists in the Milky
Way---it is only possible to specify an upper limit on its contribution to the bulge/bar
component, for which the model is still consistent with observational data, however the
data does not require the presence of a bulge (\citealt{B-HG16}).
So, we
used two models for the bulge/bar. The first, most likely at present, includes the
bar with a mass of $M_\text{bar} = 3.9 \times 10^{10}~M_{\sun}$ (\citealt{C-Dea13}) and
does not contain a bulge (hereinafter the ``only bar'' model). The second model
contains the bulge with a mass of $M_\text{bulge} = 0.78 \times 10^{10}~M_{\sun}$, which
is 20\% of the total mass of the bulge/bar 
(\citealt{B-HG16}), and the bar with $M_\text{bar} = 3.12 \times 10^{10}~M_{\sun}$.

Since we are considering motion in the close vicinity of the Galactic barycenter, we
should take into account not only large-scale components, but also the NSD, the main
component of the nuclear bulge in addition to the NSC (
\citealt{Lea02}). To represent the NSD, we also used the Miyamoto-Nagai
potential~\eqref{MNmodel} with the break radius of $90$\,pc as the parameter~$a$, the
vertical scale-height of $45$\,pc as $b$, and the mass of $M_{\text{comp}}=M_{\text{NSD}}
=  (1.4\pm0.6)\times10^9M_{\odot}$ ( \citealt{Lea02,B-HG16}). When integrating the
orbits, we applied both the point estimate of $M_{\text{NSD}}$ and values different from it
by $1\sigma$.

\section{Possible orbits of the central black hole}
\label{sect:orbits}

Orbits started from the Galactic center with the initial Galactocentric velocity in the
Galactic plane ${\mathbf V}^0(\text{BH})=(V_r^\text{LSR}(\text{BH}),V_l^0(\text{BH}))$, and
the vertical component of the initial velocity was assumed to be zero (see
\citealt{RB04}). 

\begin{figure}
   \centering
\hspace{-2.85cm}%
\includegraphics[width=0.6\textwidth, angle=0]{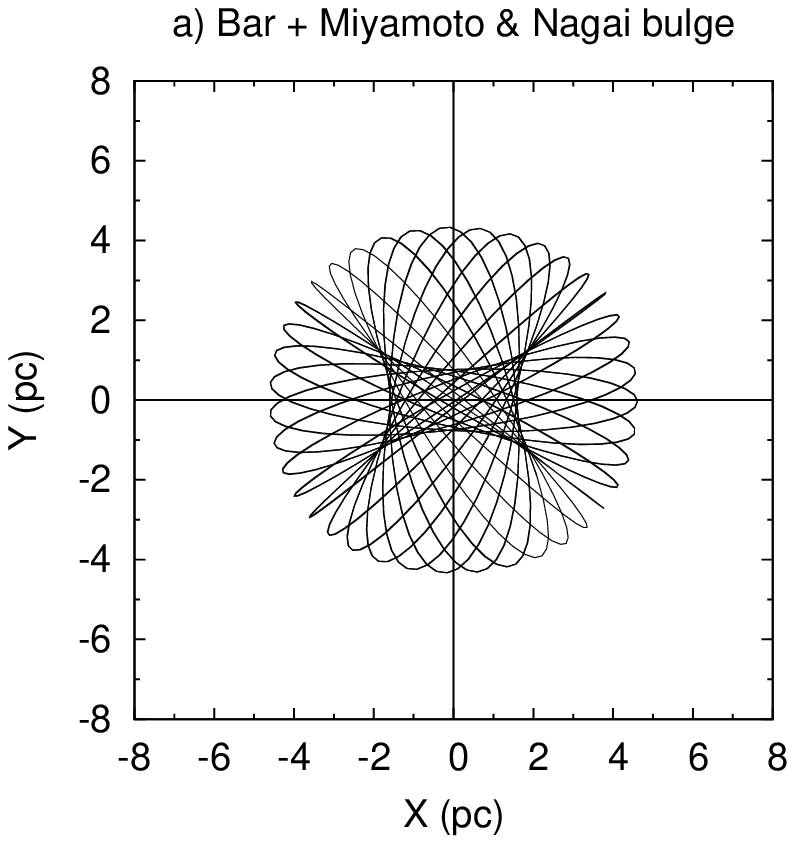}%
\hspace{-2.55cm}%
\includegraphics[width=0.6\textwidth, angle=0]{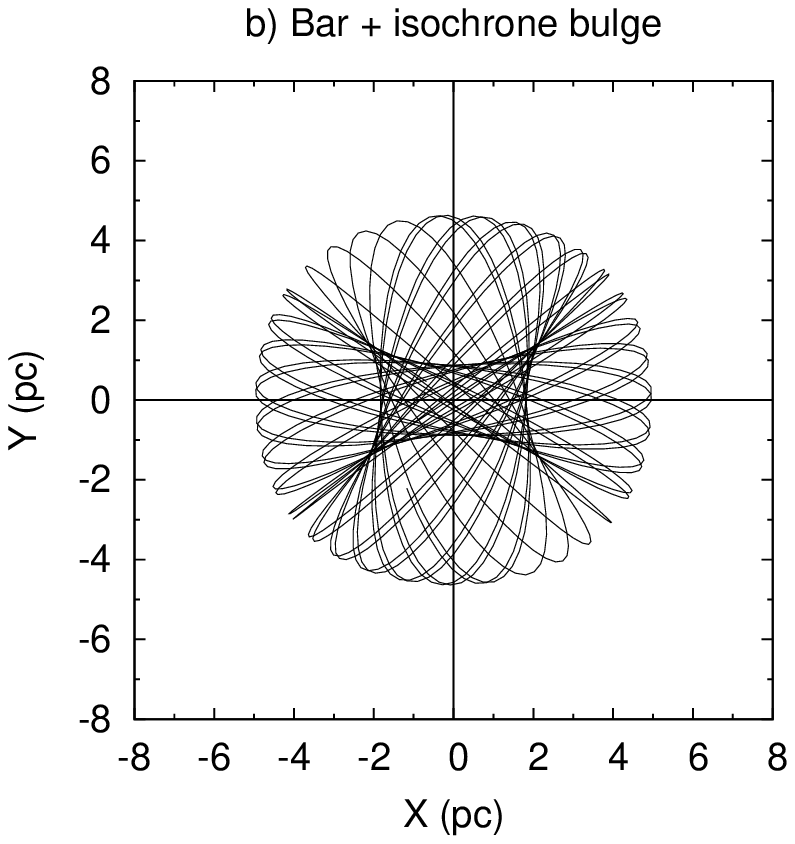}%

\hspace{-1.7cm}%
\includegraphics[width=0.6\textwidth, angle=0]{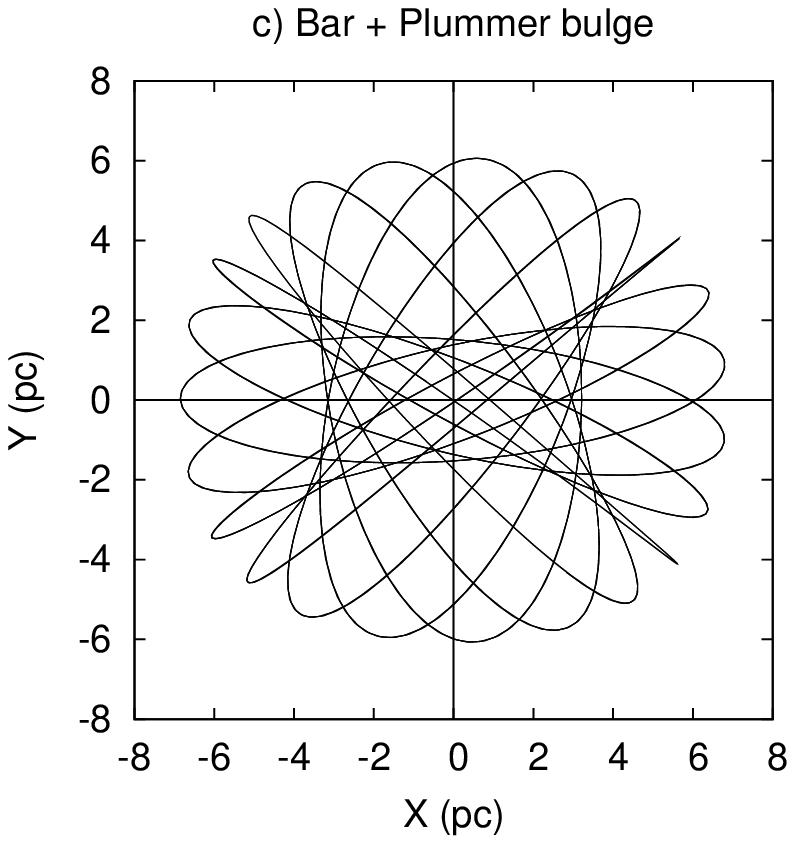}%
\hspace{-1.4cm}%
\includegraphics[width=0.6\textwidth, angle=0]{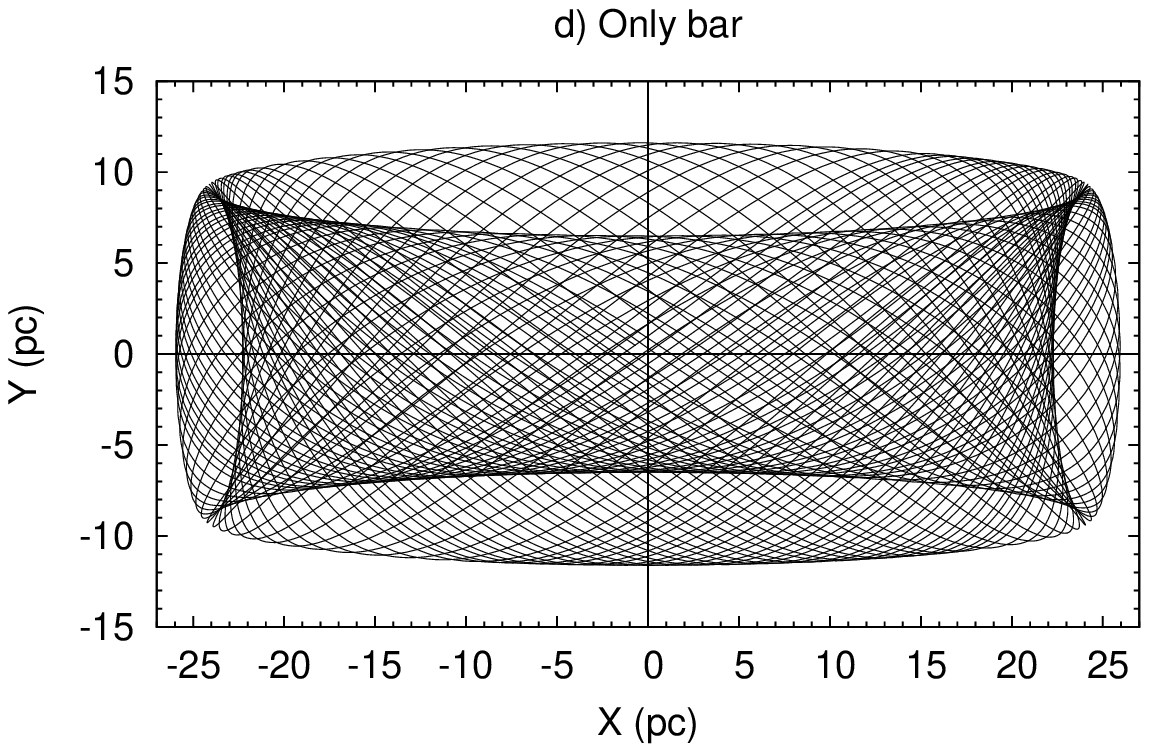}%
   \caption{Possible orbits of the SMBH for different models of the bulge/bar
 in the absence of a 
nuclear stellar disk (NSD) component in the Galactic potential model at the initial radial velocity
$V_r^\text{LSR}(\text{BH})=-4.1$\,km\,s$^{-1}$ and velocity in longitude
$V_l^0(\text{BH})=+7$\kms\ of the SMBH in the Galactic center.}
   \label{fig:nominal}
   \end{figure}

\begin{figure}
   \centering
\hspace{-2.85cm}%
\includegraphics[width=0.6\textwidth, angle=0]{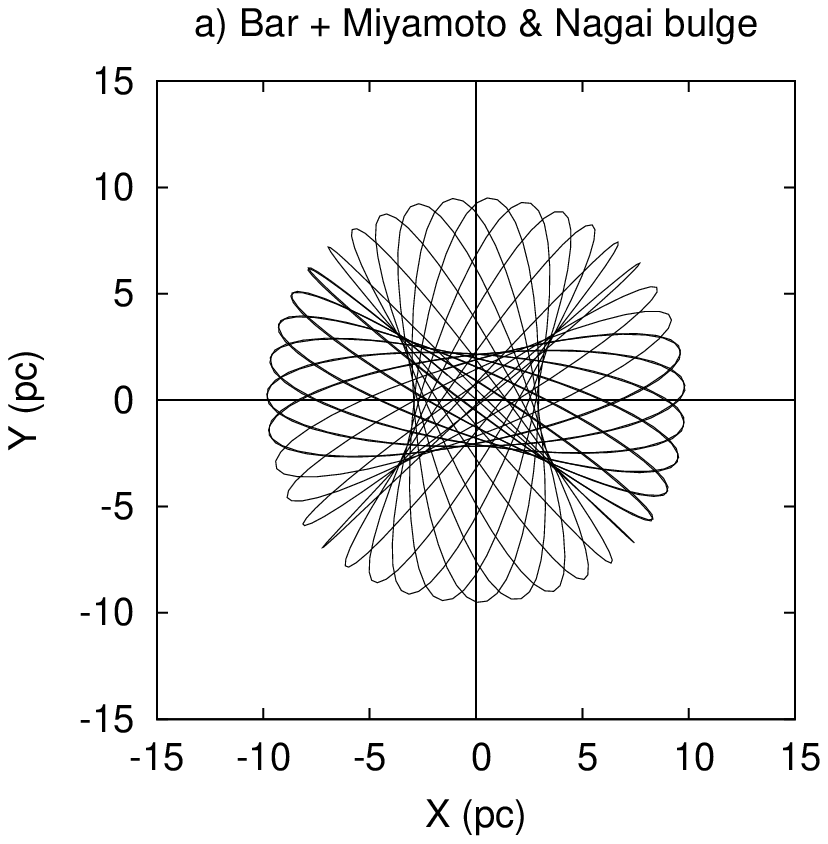}%
\hspace{-2.55cm}%
\includegraphics[width=0.6\textwidth, angle=0]{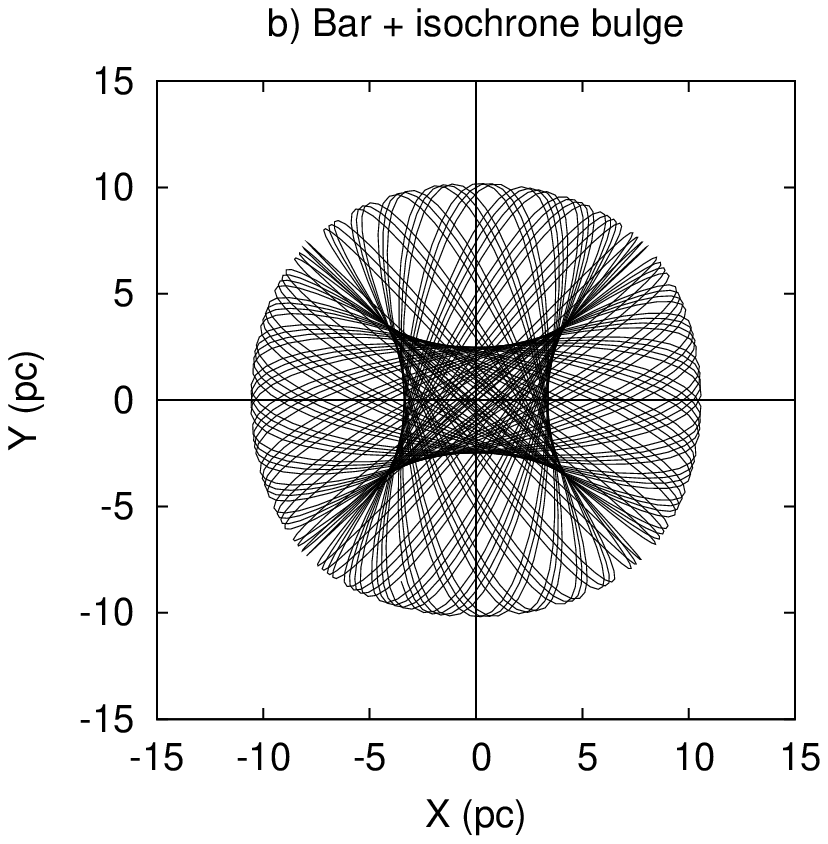}%

\hspace{-1.7cm}%
\includegraphics[width=0.6\textwidth, angle=0]{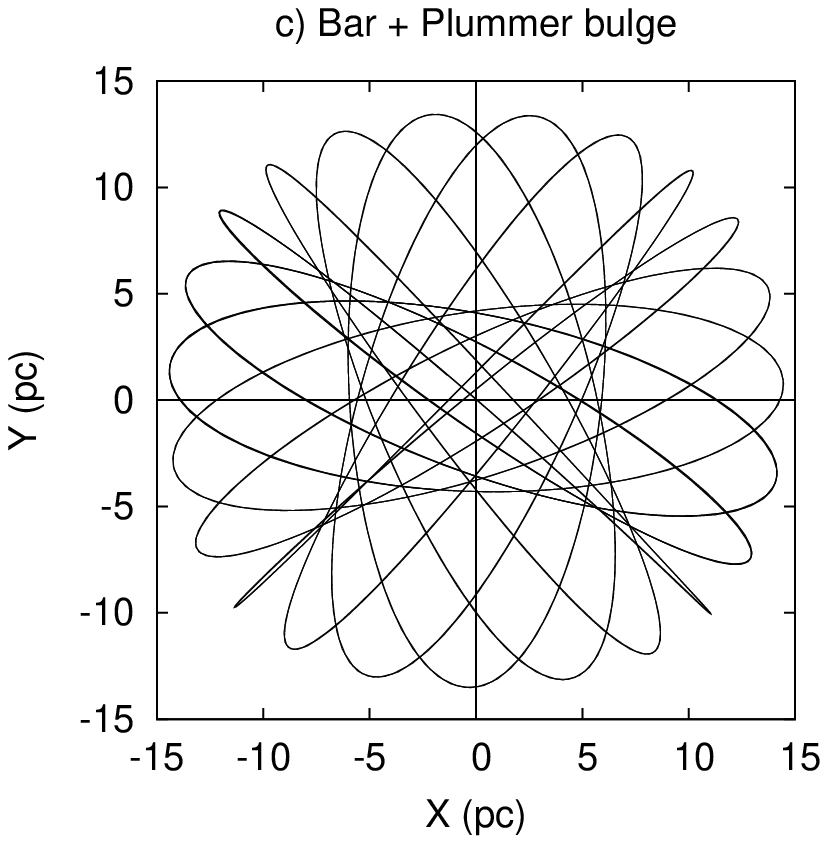}%
\hspace{-1.4cm}%
\includegraphics[width=0.6\textwidth, angle=0]{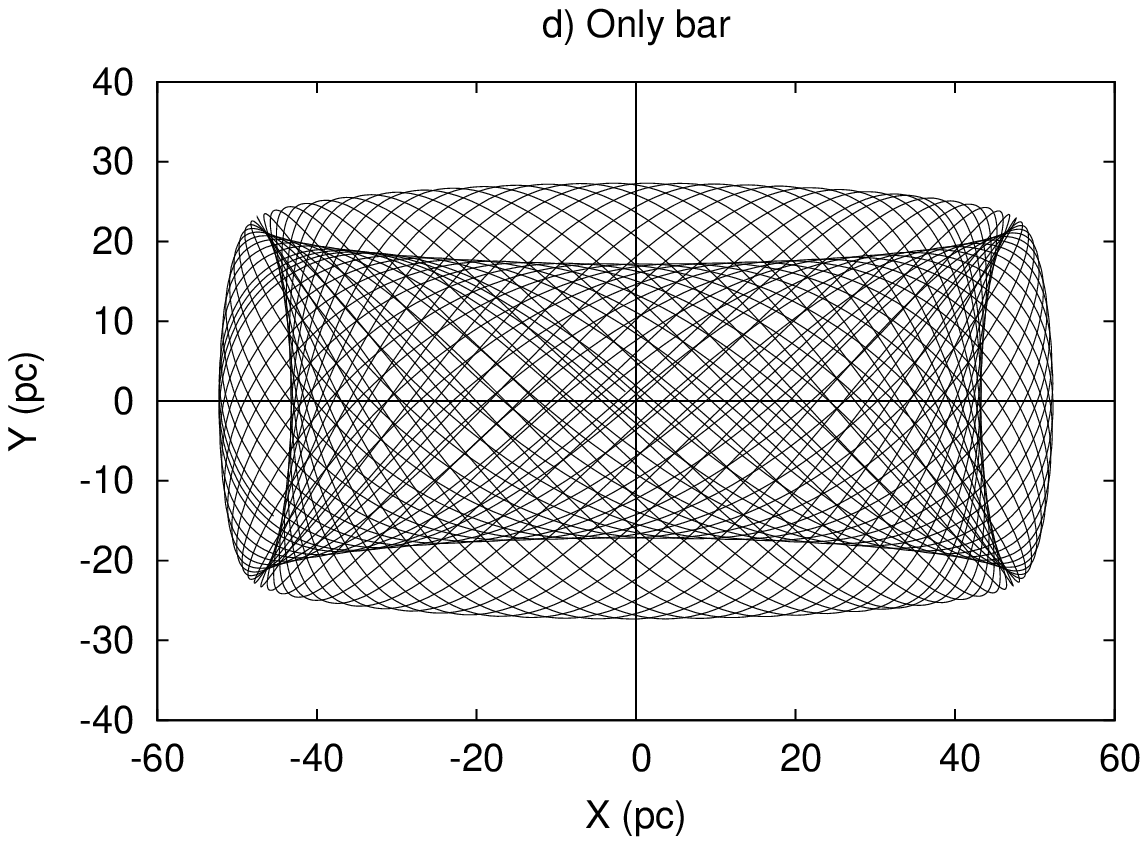}%
   \caption{As in Figure~\ref{fig:nominal}, but at the initial velocities of
$V_r^\text{LSR}(\text{BH})=-7.1$\,km\,s$^{-1}$ and $V_l^0(\text{BH})=+16$\kms.}
   \label{fig:2sigma}
   \end{figure}

In the beginning, we studied the role of the bulge/bar components.
In Figures~\ref{fig:nominal} and \ref{fig:2sigma}, we present the SMBH orbits
for the considered bulge/bar models  without taking into account the NSD component at the nominal
values of components of the velocity ${\mathbf V}^0(\text{BH})$ and at $2\sigma$-values
(see Sec.~\ref{sect:velocity}), respectively. The orbits are shown in a Galactocentric
frame of reference associated with the bar (rotating with angular velocity
$\omega_\text{bar}=40$\kmskpc). The plane $(X,Y)$ coincides with the plane of the Galaxy.
The $X$ axis is directed along the large axis of the bar, and $Y$ axis is along the small
axis. In both Figures~\ref{fig:nominal} and
\ref{fig:2sigma}, orbits for the models with classical bulge component are plotted on the
same scale, and the orbit for the non-bulge model is represented on a smaller scale.

Figures~\ref{fig:nominal} and \ref{fig:2sigma} show that the amplitude of the SMBH
oscillations relative to the barycenter of the Galaxy is not negligible in general 
for the considered models. However, the amplitude, as well as the shape of the
orbits, strongly depends on the presence of a component of the classical bulge in the
model. At the nominal peculiar/residual velocities of the SMBH the oscillation range does
not exceed~7\,pc in case of a model with bulge
(Figs.~\ref{fig:nominal}a--\ref{fig:nominal}c), but reaches~25\,pc if there is no bulge
(Fig.~\ref{fig:nominal}d). (Note that in all calculations here the total mass of the
bulge/bar  component remains constant.) At the same time, taking into
account the current uncertainty of the SMBH peculiar/residual velocity, it is impossible
to exclude the amplitude of oscillations up to $10$--$15$\,pc at the confidence level
${\approx}95\%$ even for models with bulge (Figs.~\ref{fig:2sigma}a--\ref{fig:2sigma}c).
For the ``only bar'' model  deviations of the SMBH from the
barycenter up to 50\,pc are not excluded at the significance level of $2\sigma$
(Fig.~\ref{fig:2sigma}d). We also note that the model of the Galaxy without the classical
bulge is now more reasonable (see Sect.~\ref{sect:potential}).

In the absence of a bulge, the orbits are naturally strongly elongated along the large axis
of the bar (Figs.~\ref{fig:nominal}d and \ref{fig:2sigma}d). But even with the bulge's relatively
small contribution (20\% by mass) to the bulge/bar  component, the
orbits become  almost circular envelopes
(Figs.~\ref{fig:nominal}a--\ref{fig:nominal}c,
\ref{fig:2sigma}a--\ref{fig:2sigma}c).

\begin{figure}[t!]
   \centering
\mbox{%
\hspace{-1.2cm}%
\includegraphics[width=0.65\textwidth, angle=0]{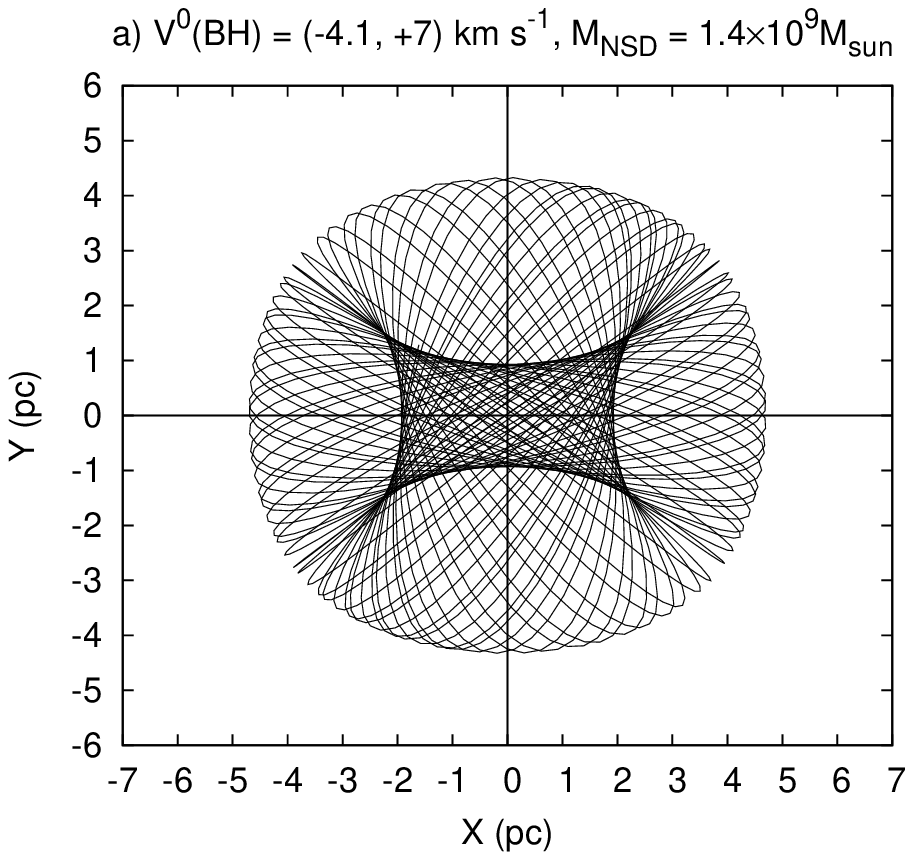}%
\hspace{-2.1cm}%
\includegraphics[width=0.65\textwidth, angle=0]{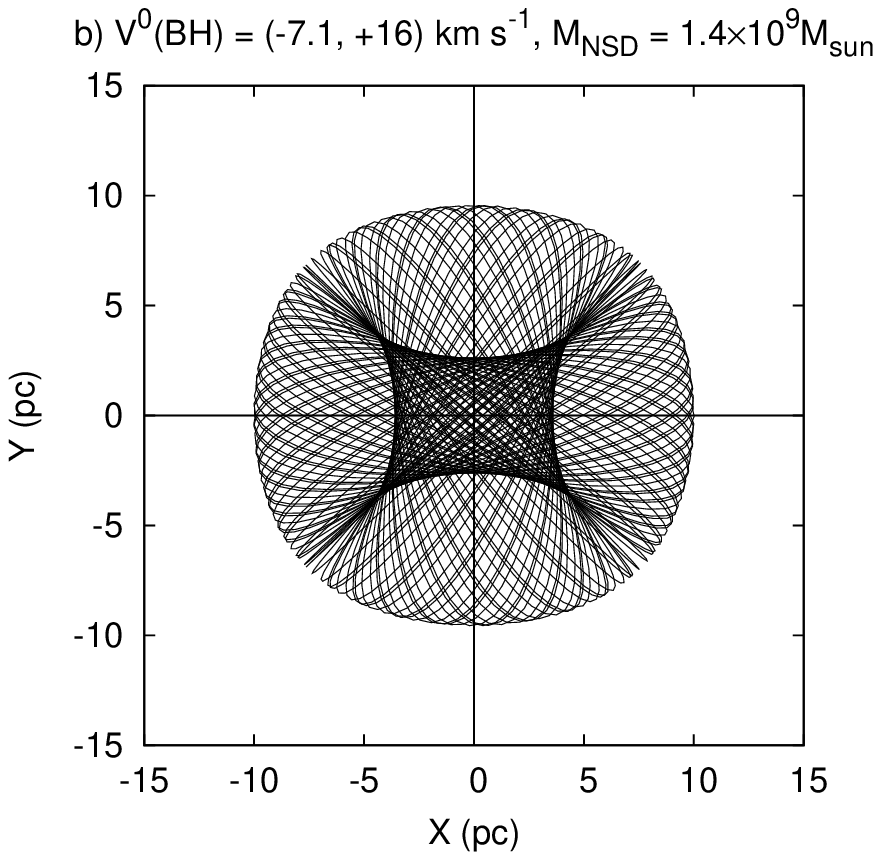}%
}%
\vspace{0.2cm}%

\mbox{%
\hspace{-1.2cm}%
\includegraphics[width=0.65\textwidth, angle=0]{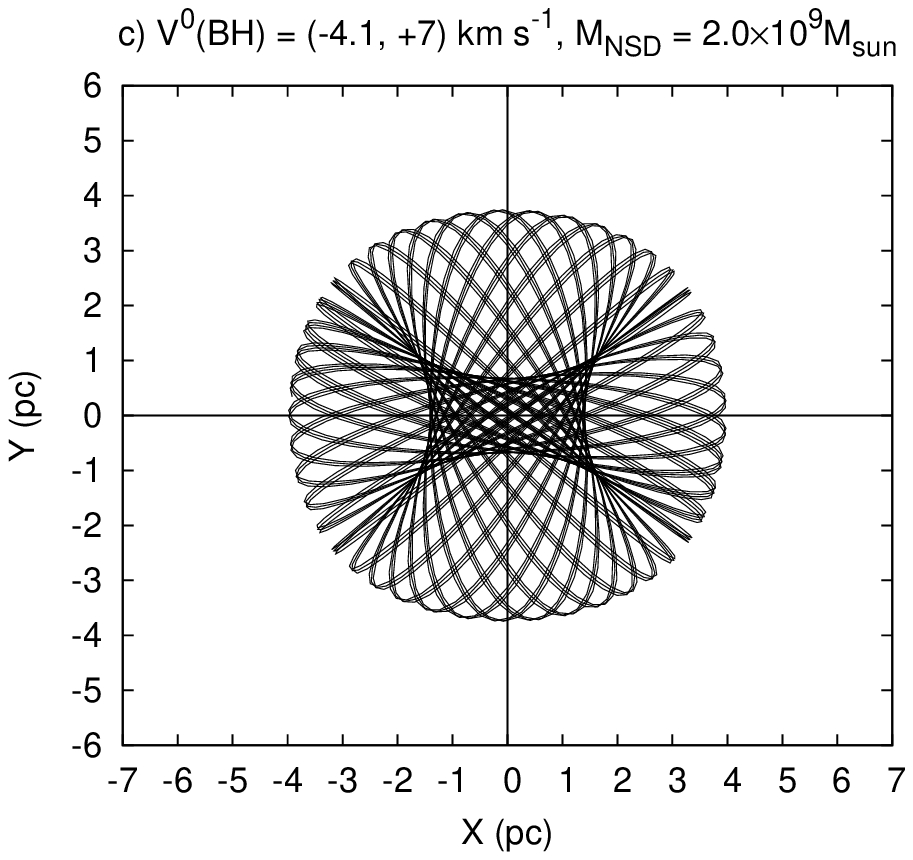}%
\hspace{-2.1cm}%
\includegraphics[width=0.65\textwidth, angle=0]{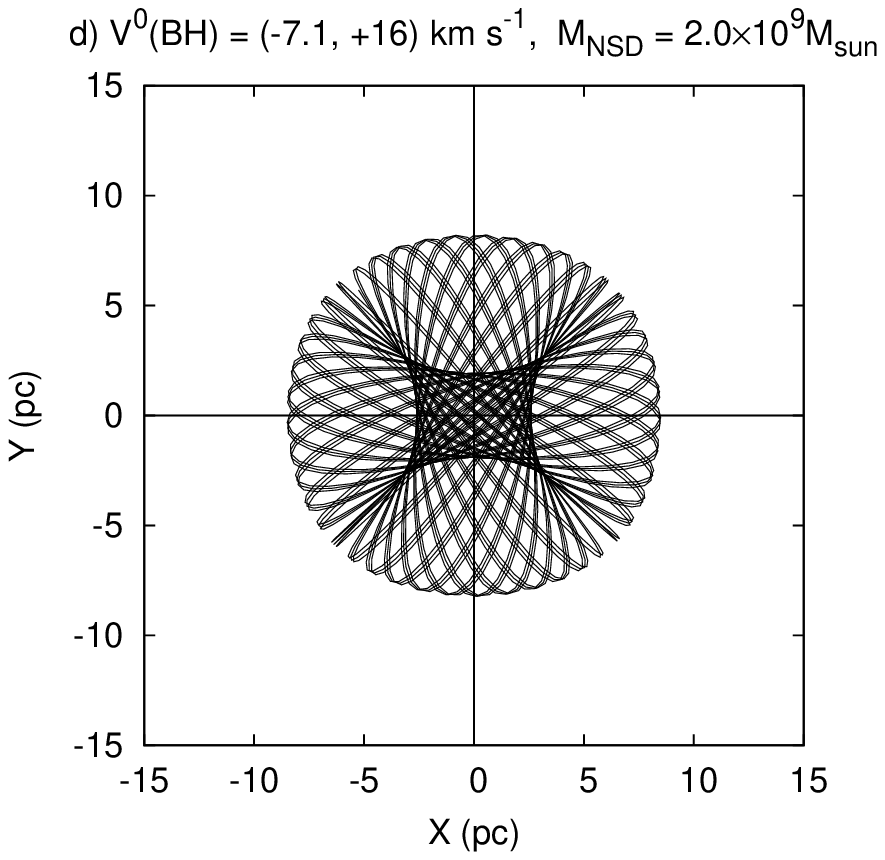}%
}%
\vspace{0.2cm}%

\mbox{%
\hspace{-1.2cm}%
\includegraphics[width=0.65\textwidth, angle=0]{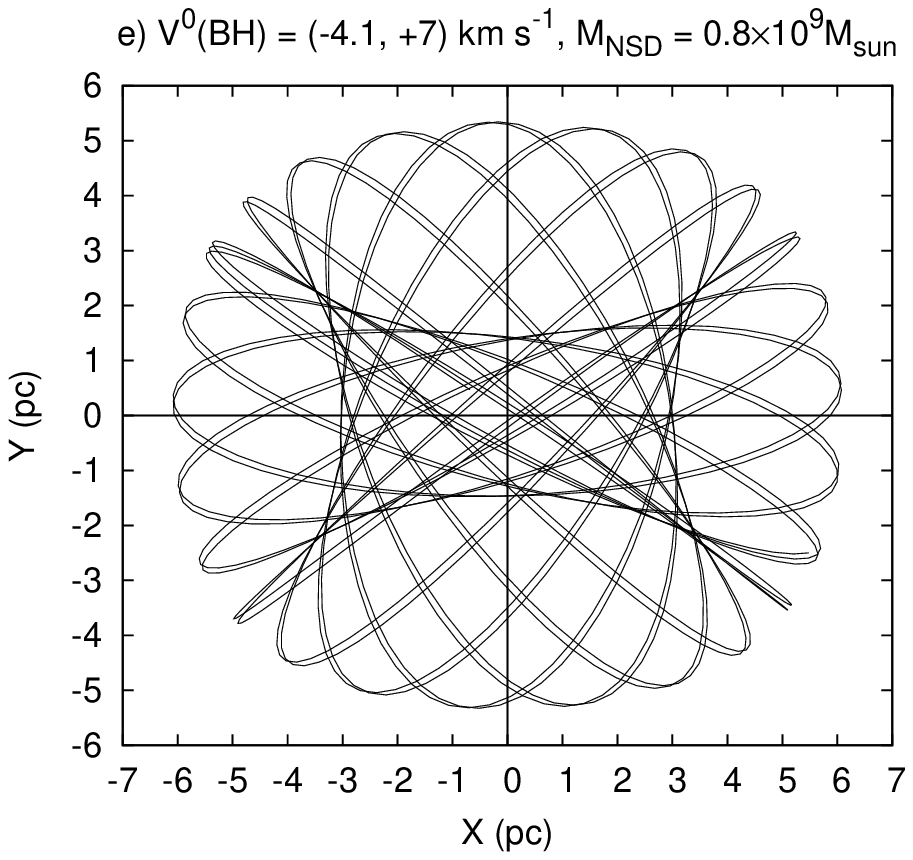}%
\hspace{-2.1cm}%
\includegraphics[width=0.65\textwidth, angle=0]{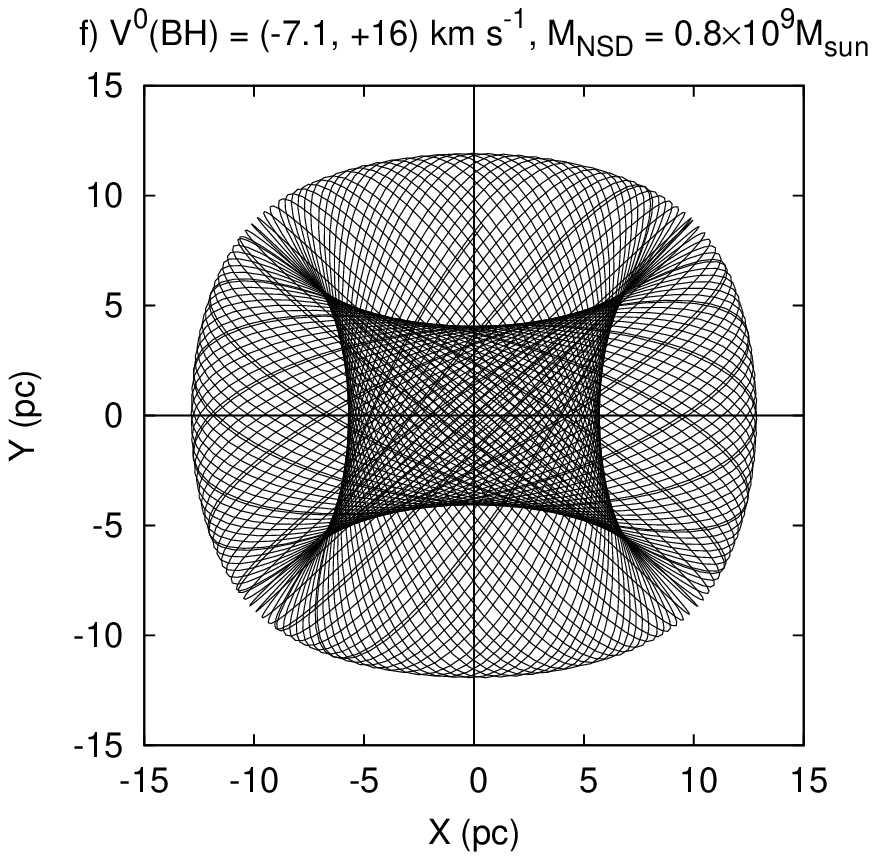}%
}%
   \caption{Possible orbits of the SMBH for the Galactic potential models, including the bar and
NSD, but not the classical bulge, at the initial SMBH's velocities
$V_r^\text{LSR}(\text{BH})=-4.1$\,km\,s$^{-1}$ and $V_l^0(\text{BH})=+7$\,km\,s$^{-1}$
(left panels) and at $V_r^\text{LSR}(\text{BH})=-7.1$\,km\,s$^{-1}$ and
$V_l^0(\text{BH})=+16$\kms (right panels) for different values of the NSD's mass:
$M_{\text{NSD}} = 1.4\times10^9M_{\odot}$ (a, b),  $2.0\times10^9M_{\odot}$ (c, d) and
$0.8\times10^9M_{\odot}$ (e, f). The scale for the right panels is smaller than that for the
left ones.}
   \label{fig:NSD}
   \end{figure}

Then we excluded the classical bulge from the model potential as an unconfirmed
component of the Galaxy, but added the NSD to the bar (``bar~+ NSD'' model), preserving
the previous value of the total mass of the central components ($M_\text{bar+NSD} = 3.9
\times 10^{10}~M_{\sun}$) in all variants. The orbits obtained for this model with a point
estimate of the NSD's mass $M_{\text{NSD}} = 1.4\times10^9M_{\odot}$ found by
\cite{Lea02} are plotted in Figures~\ref{fig:NSD}a and \ref{fig:NSD}b. The picture of the
orbital motions has changed dramatically: the orbits have turned out to be much more
compact and more rounded than for the "bar only" model (cf. Figs.~\ref{fig:nominal}d,
\ref{fig:2sigma}d), and have become close to those obtained with the presence of the bulge
component (Figs.~\ref{fig:nominal}a--\ref{fig:nominal}c,
\ref{fig:2sigma}a--\ref{fig:2sigma}c). At the nominal value of velocity ${\mathbf
V}^0(\text{BH})$, the oscillation range is only 4.7\,pc, and at the $2\sigma$-value it is
10\,pc.

Variation in the mass of the NSD by $\pm1\sigma$ ($\pm0.6\times10^9M_{\odot}$, see 
\citealt{Lea02}), although it gives an asymmetric effect, namely,
the increase in mass leads to a reduction in the oscillation range by 10--15\%; the decrease
leads to an increase in the range by 26--34\% (Figs.~\ref{fig:NSD}c--\ref{fig:NSD}f),
but does not significantly change the results---the oscillations remain quite limited
(within 13\,pc).

\section{Discussion}
\label{sect:discussion}

Thus, of the central components of the Galactic potential confirmed by observations
and significant in mass, the NSD, although it has a shallow mass
distribution (\citealt{Lea02}), most strongly restricts the possible movement of the
SMBH\,+\,NSC complex in the regular Galactic field. Even at the highest velocity ${\mathbf
V}^0(\text{BH})$ and the lowest mass $M_{\text{NSD}}$, the considered oscillations of
the complex do not go beyond $13\text{\,pc}$. However, with the current accuracy of
measuring these parameters, it is impossible to exclude oscillations of the SMBH\,+\,NSC of
the specified and larger scale (the formal probability of finding the orbit within 13\,pc
is only 65\%, and the high uncertainty of the mass $M_{\text{NSD}}$ does not allow
performing $M_{\text{NSD}}$ variations within a large range, remaining within this
statement of the problem). A possible deviation of ${\sim}13\,\text{pc} = 6'$ (at
$R_0\sim8$\,kpc) is small compared to the size of the NSD,
${\sim}200$--$400\,\text{pc}=1\fdg5$--$3\dg$, and modern infrared images and stellar
number density maps of the nuclear bulge (e.g.,
\citealt{Nea13,G-Cea20}) do not exclude it.

\begin{figure}
   \centering
\includegraphics[width=0.8\textwidth, angle=0]{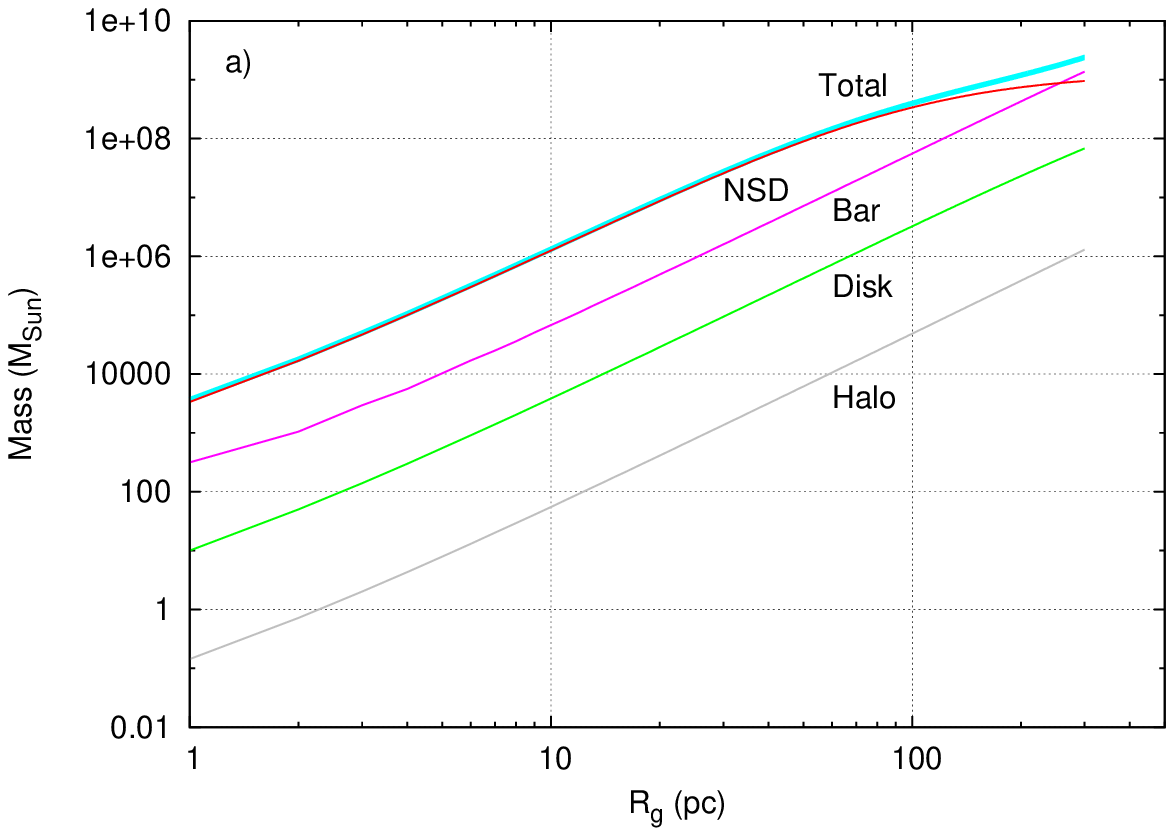}%

\includegraphics[width=0.8\textwidth, angle=0]{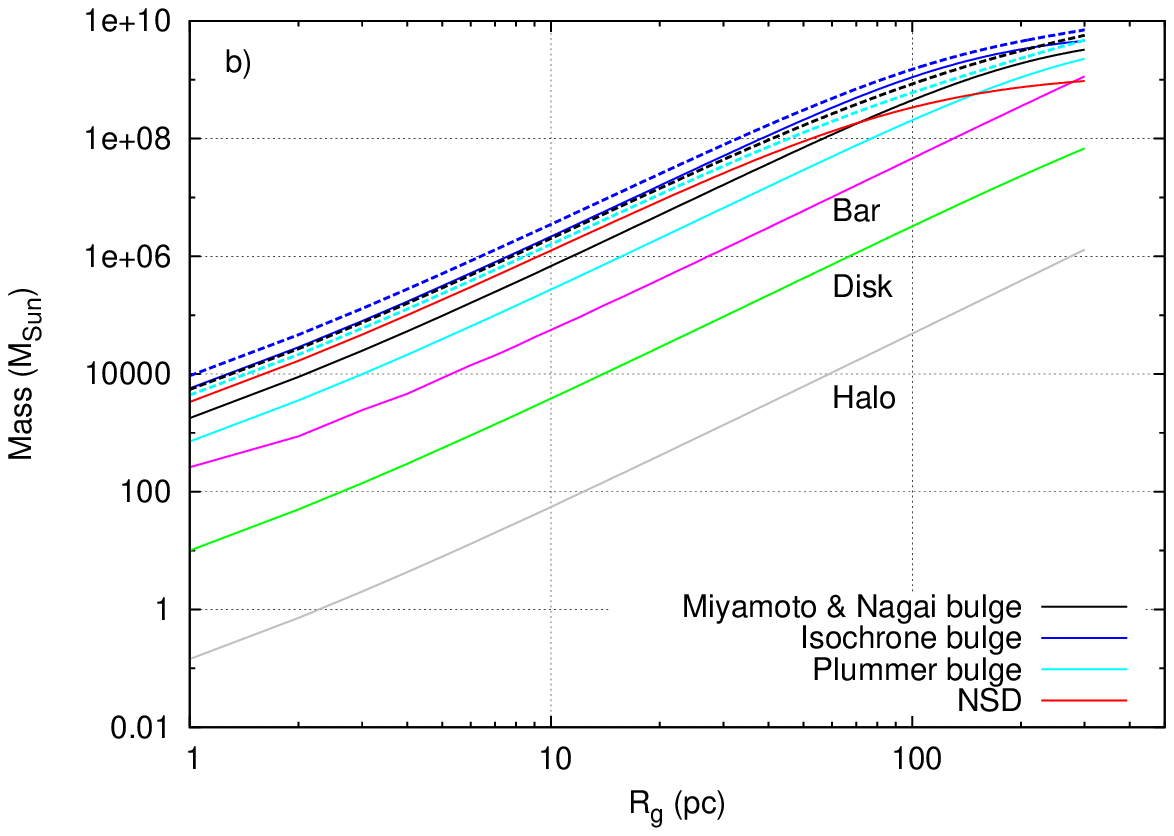}%
   \caption{Mass enclosed in spheres of radius $R_\text{g}$ 
for the ``bar~+ NSD'' model~(a) and when adding a classical bulge component to this model~(b).
The full mass profiles in panel (b) are shown as dashed lines; the color of each of these
profiles coincides with the color of the contributing profile of the corresponding bulge
model.}
   \label{fig:mass profiles}
   \end{figure}

The important role of the NSD in this problem is not surprising, since it dominates the
area of possible movements of the SMBH\,+\,NSC under the accepted ``bar~+ NSD'' model
(Fig.~\ref{fig:mass profiles}a). It is interesting that adding the classical bulge to this
model ($M_\text{NSD+bulge} = 0.78 \times 10^{10}~M_{\sun}$, $M_\text{bar} = 3.12 \times
10^{10}~M_{\sun}$) leads to an extra mass in the outer region of the NSD
(Fig.~\ref{fig:mass profiles}b) compared to the mass profile constructed by \cite{Lea02} 
(cf. fig.~14 in their work). For example, the $10^8~M_{\sun}$ level is reached at a radius
of $R_\text{g}=31$--43\,pc when adding a bulge, and for the ``bar~+ NSD'' model at
$R_\text{g}=50$\,pc, as on the \cite{Lea02}   profile (the corresponding level there is
$1.3\times10^8~M_{\sun}$, since the masses of the SMBH and NSC were taken into account
when building the profile). That is, when accounting for the NSD, additional introduction
of the classical bulge into the model now seems redundant, at least on the scale of the
nuclear bulge, in agreement with the conclusions regarding the bulge in \cite{B-HG16}. 

However, despite the influence of the NSD, the effect of potential asymmetry is
noticeable: the orbits in Figure~\ref{fig:NSD} are slightly elongated along the large axis
of the bar. This is especially evident in the contours of the direction field that
restrict regions with a four-fold field. Note that all the orbits constructed in this
paper have the latter regions.

We tested the possible scale of deviations of the SMBH from the center of the NSC
using a ``naive'' model, adding an NSC component to the ``bar + NSD'' model. Since the
cluster shape is not spherically symmetric---the axis ratio is $c/a = 0.71 \pm 0.04$ (
\citealt{B-HG16})---we did not consider the Plummer potential model, but adopted the
Miyamoto \& Nagai model~\eqref{MNmodel} with parameter values of the half-light radius
$a=4.2$\,pc, $b = 3$\,pc and $M_{\text{NSC}} = 1.8\times10^7M_{\odot}$
(\citealt{B-HG16}). Even with the specified moderate mass of the NSC, at the nominal
${\mathbf V}^0(\text{BH})$ the deviations do not exceed 0.6\,pc, which
can be ignored at the present stage of the analysis.

It should be noted that possible oscillations of the gravitationally bound SMBH\,+\,NSC
complex should hardly prevent the formation of a cuspy stellar distribution detected near
\sgra (e.g. {\citealt{G-Cea20}). Massive stars of the nuclear bulge are concentrated
not only in the NSC, but also in several other clusters, including Sgr B2, Sgr B1, Sgr C
and others ({\citealt{Lea02}). This means that the movement of these concentrations
in the gravitational field does not interfere with star formation in them. A stellar cusp
is then formed in the concentration in which there is the SMBH ({\citealt{Baumgardt+18}),
i.e. in NSC.

Note that the non-central start of the orbits increases the oscillation range. Taking into
account irregular forces should lead to stochastization of motion, i.e., in general to
 greater deviations from the center. In this sense,  the
oscillation amplitudes obtained here are  estimates from below.
However, if a bulge component is still detected, it will, on the contrary, lead to
stabilization of oscillations.

The obtained results suggest that at present it is impossible to exclude the
non-centrality of position of the SMBH (\sgra) with a deviation from the Galactic
barycenter only on the scale of about a dozen parsecs. The marginal significance of the SMBH
peculiar radial velocity $V_r^\text{LSR}(\text{BH})$ (\citealt{GC19}) supports this
possibility. The specified scale of SMBH deviation is still
insignificant compared to the measurement precision and accuracy of $R_0$.

\begin{acknowledgements}
We thank the anonymous referee for a very helpful report.
I.I.N.\ acknowledges support from the Russian Foundation for Basic Research, grant
no.~18-02-00552.

\end{acknowledgements}

\label{lastpage}

\end{document}